\documentclass{llncs}

\usepackage[T1]{fontenc}
\usepackage{graphicx}
\usepackage{booktabs}
\usepackage{multirow}
\usepackage{tabularx}
\usepackage{amsmath,amssymb}
\usepackage{microtype}
\usepackage{url}
\usepackage{xcolor}
\usepackage{xspace}
\usepackage{flafter}
\usepackage{float}
\usepackage{placeins}

\newcommand{\cahnm}{\textsc{CA-HNM}\xspace}
\newcommand{\cahnpure}{\textsc{CA-HNM}-Pure\xspace}
\newcommand{\cahnmix}{\textsc{CA-HNM}-Mixed\xspace}
\newcommand{\ndcg}{NDCG@10\xspace}
\newcommand{\mrr}{MRR@10\xspace}
\newcommand{\mapscore}{MAP@100\xspace}
\newcommand{\recall}{Recall@100\xspace}

\makeatletter
\renewcommand\section{\@startsection{section}{1}{\z@}%
	{0.8ex \@plus 0.8ex \@minus 0.8ex}%
	{0.8ex \@plus 0.8ex}%
	{\normalfont\large\bfseries}}
\renewcommand\subsection{\@startsection{subsection}{2}{\z@}%
	{0.7ex \@plus 0.7ex \@minus 0.7ex}%
	{0.7ex \@plus 0.7ex}%
	{\normalfont\normalsize\bfseries}}
\renewcommand\subsubsection{\@startsection{subsubsection}{3}{\z@}%
	{0.7ex \@plus 0.7ex \@minus 0.7ex}%
	{0.7ex \@plus 0.7ex}%
	{\normalfont\normalsize\itshape}}
\makeatother

\begin{document}
\raggedbottom

\title{\textsc{CA-HNM}: Constraint-Aware Hard Negative Mining for Dense
Retrieval with Structured Domain Models}

\author{
	Ng\d{o}c Luy\d{\^{e}}n L\^e\inst{1,2}\orcidID{0000-0002-2043-0679}
	\and
	Marie-H\'el\`ene Abel\inst{2}\orcidID{0000-0003-1812-6763}
	\and
	Bertrand Laforge\inst{3}\orcidID{0000-0002-4209-4194}
}

\authorrunning{N. L. L\^e et al.}

\institute{
	Gamaizer, 93340 Le Raincy, France
	\and
	Universit\'e de technologie de Compi\`egne, CNRS, Heudiasyc
	(Heuristics and Diagnosis of Complex Systems),
	CS 60319, 60203 Compi\`egne Cedex, France\\
	\email{\{ngoc-luyen.le,marie-helene.abel\}@hds.utc.fr}
	\and
	Sorbonne Universit\'e, CNRS UMR 7585, LPNHE
	(Laboratoire de Physique Nucl\'eaire et des Hautes \'Energies),
	75252 Paris Cedex 05, France\\
	\email{laforge@lpnhe.in2p3.fr}
}

\maketitle

\begin{abstract}
Dense retrievers are commonly trained with hard negatives selected from lexical
or dense retrieval results. Retrieval proximity, however, does not indicate
whether a candidate conflicts with explicit domain structure. We introduce
\cahnm, a constraint-aware hard-negative mining framework that combines
retrieval-based candidate generation with deterministic incompatibility rules
derived from a structured domain model. The framework removes known positives,
constructs a local typed context around the query target, records relational,
scope, target, and level mismatches, and uses the resulting candidates to
fine-tune a standard dual encoder. Structured information is required only during
offline negative construction. We evaluate \cahnm on MOOCCubeX and Course-Skill
Atlas with two encoders, two training losses, and ten seeds. Relative to a
retrieval-rank-matched control, \cahnpure has higher mean \ndcg in six of eight
dataset--encoder--loss configurations. Four hierarchical 95\% bootstrap intervals are entirely above zero, mostly
under triplet loss. In a separate experiment, \cahnmix shows slightly higher mean \ndcg
than a matched-mixed control on both collections. Structural variants show
that changes in available structure can also change candidate eligibility and
negative-pool composition. The results indicate that the contribution of
constraint-aware selection depends on the training objective and the composition
of the negative pool.

\end{abstract}

\section{Introduction}
\label{sec:introduction}

Dense retrieval models are typically trained by contrasting a relevant document
with one or more documents treated as negatives. The choice of negatives has a direct
effect on the training signal: random negatives are often too easy, whereas
lexical or dense retrieval can provide candidates that are close to the query and
therefore more informative~\cite{karpukhin2020dense,xiong2020approximate,qu2021rocketqa,zhan2021optimizing}.
Existing hard-negative strategies consequently define hardness mainly through
lexical score, embedding similarity, or retrieval rank. These signals are useful
for locating difficult candidates, but they do not determine whether an unjudged
candidate is a valid negative. Incomplete relevance judgments can leave missing
positives or ambiguous cases among highly ranked documents~\cite{cai2022hard}.

Structured domains provide another source of information. Concepts may be linked
through prerequisite, hierarchical, related-concept, scope, or level relations.
A document can therefore be retrieval-hard while still conflicting with the
target represented by the query. For example, a resource about advanced SQL
optimization may be close to a query for ``beginner SQL joins'' even though its
target concept and level differ. Such structure can be used during training-data
construction without changing the retriever architecture.

We introduce \cahnm, a constraint-aware hard-negative mining framework. Starting
from lexical and dense retrieval results, \cahnm removes known positives,
constructs a local typed context for the query target, evaluates the remaining
candidates with deterministic compatibility rules, and records the rule evidence
associated with each selected negative. The structured model is used only during
offline negative construction; training and inference use a standard dual
encoder.

To evaluate the proposed selection mechanism, the empirical study addresses
three research questions: \textbf{RQ1} - How does constraint-aware negative selection compare with
  retrieval-rank-matched selection? \textbf{RQ2} - What is the contribution of constraint-aware negatives
  within a mixed negative pool?
  \textbf{RQ3} - How do changes in the structured information available to
  the classifier affect negative selection and retrieval performance?
The evaluation covers MOOCCubeX and Course-Skill Atlas, BGE and E5 encoders~\cite{xiao2023cpack,wang2022textembeddings},
triplet loss and cached multiple-negatives ranking loss (cached MNRL)~\cite{gao2021scaling}, and ten training seeds.

The contributions are threefold. First, we formulate hard-negative mining by
separating retrieval hardness, positive annotation, and operational
incompatibility under a structured domain model. Second, we define a mining
pipeline that associates selected negatives with explicit relation, scope,
target, or level evidence while leaving the inference-time retriever unchanged.
Third, we provide a multi-seed evaluation aligned with these research
questions through rank-matched, matched-mixed, and structural controls across
multiple encoders, losses, and datasets.

The remainder of the paper is organized as follows.
Section~\ref{sec:related} reviews hard-negative mining, generated supervision,
and structured knowledge in retrieval. Section~\ref{sec:problem} defines the
problem, and Section~\ref{sec:method} presents \cahnm.
Section~\ref{sec:setup} describes the experimental protocol.
Section~\ref{sec:results} reports the results, followed by the discussion in
Section~\ref{sec:discussion} and the conclusion in
Section~\ref{sec:conclusion}.

\section{Related Work}
\label{sec:related}

Prior work relevant to this study spans three areas: hard-negative mining for
dense retrieval, synthetic retrieval supervision, and the use of structured
knowledge in information retrieval. We focus on how these approaches construct
or select training signals for retrieval models.

%\subsection{Dense Retrieval and Hard-Negative Mining}

Dense Passage Retrieval and Sentence-BERT use bi-encoder representations for
efficient semantic retrieval~\cite{karpukhin2020dense,reimers2019sentence},
while ColBERT and SPLADE use late interaction and learned sparse
representations~\cite{khattab2020colbert,formal2021splade}. BEIR highlights
performance variation across domains and tasks~\cite{thakur2021beir}.

For negative selection, ANCE mines approximate nearest
neighbors~\cite{xiong2020approximate}, RocketQA combines large candidate pools
with denoising~\cite{qu2021rocketqa}, ADORE selects highly ranked dense
negatives~\cite{zhan2021optimizing}, and BM25 supplies lexically similar
candidates~\cite{robertson2009probabilistic}. Negative effectiveness depends
on difficulty and sampling strategy~\cite{hofstaetter2021efficiently,rajapakse2024negative}.
However, unjudged high-ranking documents may be relevant when judgments are
incomplete~\cite{cai2022hard}, motivating selection signals beyond retrieval
proximity.

%\subsection{Synthetic and Generated Retrieval Supervision}

A complementary line of work creates additional supervision for retriever
training. Promptagator generates task-specific queries from a small number of
examples~\cite{dai2022promptagator}, InPars generates synthetic
query--document pairs~\cite{bonifacio2022inpars}, and GPL combines query
generation with pseudo-labeling for domain adaptation
~\cite{wang2022gpl}. SyNeg more directly generates semantic hard negatives
using large language models~\cite{li2024syneg}.
These approaches modify or generate training examples from text or model
predictions. They demonstrate that retriever supervision can be improved by
controlling the composition of the training data, but they do not generally
exploit explicit domain relations to determine whether an already retrieved
candidate is compatible with the query target.

%\subsection{Structured Knowledge in Information Retrieval}

Ontologies, taxonomies, and knowledge graphs have been used for semantic
search, query expansion, entity-oriented retrieval, and explanation
~\cite{vallet2005ontology,reinanda2020knowledge}.
Explicit constraints over RDF knowledge graphs have also been
used to identify recommendations compatible with user
preferences~\cite{le2023constraint}. Structured knowledge is also
used in knowledge-intensive retrieval and retrieval-augmented generation, where
retrieved evidence supports downstream reasoning or generation
~\cite{petroni2021kilt,lewis2020retrieval}. In these settings, structured
information is typically incorporated during retrieval, reranking, reasoning,
or generation.
Educational retrieval provides a suitable setting for studying structured
signals because learning resources are often associated with explicit
concept relations.
Competency graphs have also been used to guide learning-resource
annotation with supporting textual
evidence~\cite{le2026learning}. MOOCCubeX provides learning concepts and prerequisite relations
~\cite{yu2021mooccubex}, while Course-Skill Atlas provides hierarchical and
related-skill structure over higher-education curricula
~\cite{javadian2024course}.

Taken together, prior work has studied retrieval-based hardness, generated
supervision, and structured knowledge mainly as separate mechanisms. The gap
addressed in this paper is the use of structured-domain incompatibility
specifically during offline hard-negative selection. \cahnm follows this
direction while leaving the underlying dual-encoder architecture and
inference procedure unchanged.

\section{Problem Formulation}
\label{sec:problem}

We distinguish three properties involved in hard-negative construction:
retrieval hardness, absence of positive annotation, and incompatibility under an
explicit structured domain model.

\subsection{Dense Retrieval and Retrieval Hardness}
\vspace{0mm}

Let $\mathcal{Q}$ be a set of queries and $\mathcal{D}$ a collection of
documents. For each query $q\in\mathcal{Q}$, let
$P(q)\subseteq\mathcal{D}$ denote the set of documents annotated as relevant.
Because relevance judgments may be incomplete, $d\notin P(q)$ means that $d$ is
not a known positive; it does not by itself establish non-relevance.

A dual encoder with parameters $\theta$ maps queries and documents into a shared
vector space and computes
\begin{equation}
  s_\theta(q,d)=\langle f_\theta(q),g_\theta(d)\rangle .
  \label{eq:retrieval-score}
\end{equation}
For a positive document $d^{+}\in P(q)$ and a sampled negative set
$\mathcal{N}(q)$, the training objective depends on the construction of
$\mathcal{N}(q)$. Conventional hard-negative mining selects unjudged candidates
that satisfy a retrieval-hardness criterion:
\begin{equation}
  \mathcal{N}_{\mathrm{ret}}(q)=
  \{d\in\mathcal{D}\setminus P(q):H(q,d)\geq\lambda\}.
  \label{eq:retrieval-hard}
\end{equation}
Here, $H(q,d)$ is defined so that larger values correspond to greater hardness;
it may be based on BM25 score, dense similarity, or an equivalent transformation
of retrieval rank. Equation~\ref{eq:retrieval-hard} identifies difficult
candidates but does not provide a domain-based incompatibility criterion.

\subsection{Structured Incompatibility}
\vspace{0mm}

We represent the structured information used by the miner as
\begin{equation}
  \mathcal{O}=(\mathcal{C},\mathcal{R},\mathcal{A},\mathcal{M}),
  \label{eq:domain-model}
\end{equation}
where $\mathcal{C}$ is a set of concepts, $\mathcal{R}$ is a set of typed
relations, $\mathcal{A}$ contains concept attributes such as labels, aliases,
descriptions, or levels, and
$\mathcal{M}\subseteq\mathcal{D}\times\mathcal{C}$ contains document--concept
associations. The structured component may correspond to an ontology, taxonomy,
knowledge graph, controlled vocabulary, or typed concept graph. For each query
$q$, let $c_q\in\mathcal{C}$ denote its target concept.

The structured model is not assumed to provide complete relevance judgments.
Instead, we define an operational incompatibility indicator
$I(q,d,\mathcal{O})$. It is one when a candidate satisfies at least one
implemented incompatibility rule and zero otherwise. The associated violation
set is
\begin{equation}
  \phi(q,d,\mathcal{O})\subseteq\mathcal{T},
  \label{eq:violation-set}
\end{equation}
where $\mathcal{T}$ contains relation, scope or granularity, target, and level
mismatch types. The violation set records the rule evidence used during mining;
it is not an independent relevance judgment.

\begin{figure*}[h]
  \centering
  \includegraphics[width=.99\textwidth]{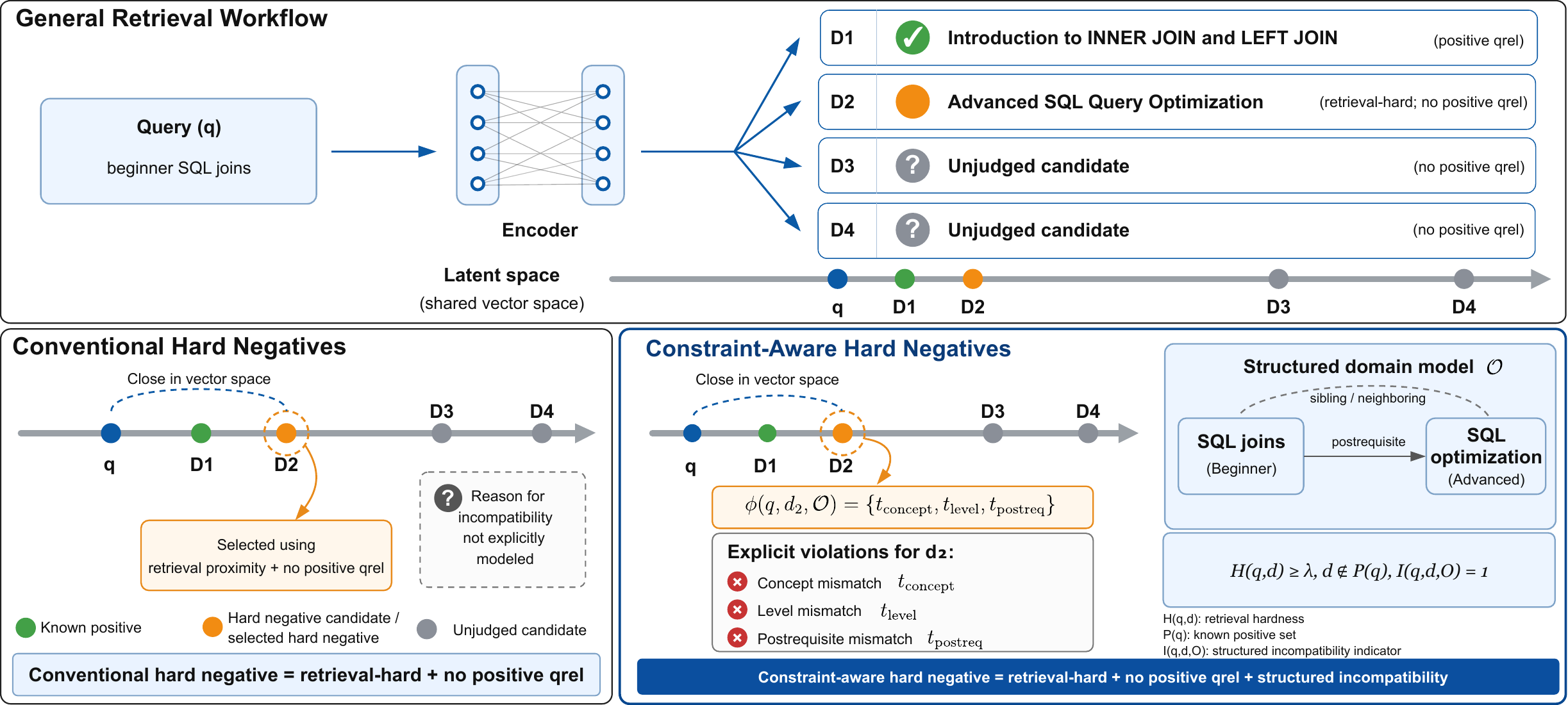}
  \caption{Conventional versus constraint-aware hard-negative selection.
  Retrieval identifies candidates close to the query, while constraint-aware
  mining additionally evaluates their compatibility with the target concept and
  its local structured context.}
  \label{fig:running-example}
\end{figure*}

Figure~\ref{fig:running-example} illustrates the distinction. Retrieval can place
a candidate close to the query without specifying why that candidate is
inappropriate. The structured context provides an additional signal through
concept relations, scope, target evidence, or level information.

A constraint-aware hard negative is therefore a retrieval-hard candidate that is
not a known positive and satisfies at least one incompatibility rule:
\begin{equation}
  \mathcal{N}_{\mathcal{O}}(q)=
  \{d\in\mathcal{D}\setminus P(q):
  H(q,d)\geq\lambda \land I(q,d,\mathcal{O})=1\}.
  \label{eq:constraint-negative}
\end{equation}
The empirical question is whether selecting such candidates changes retriever
training relative to candidates drawn from comparable retrieval positions without
consulting the structured model.

\section{Constraint-Aware Hard Negative Mining}
\label{sec:method}

\begin{figure*}[t]
  \centering
  \includegraphics[width=.99\textwidth]{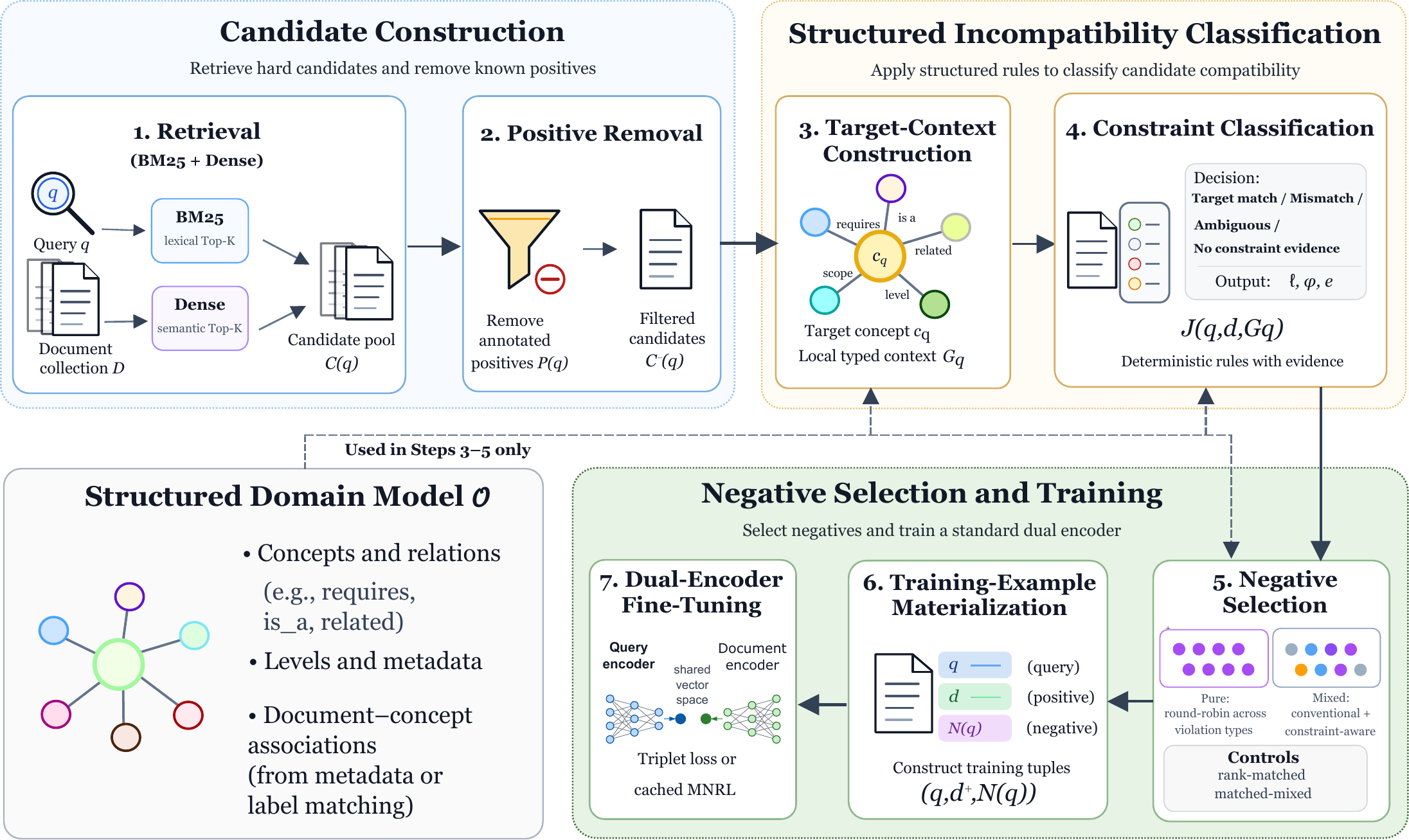}
  \caption{Overview of \cahnm. Lexical and dense retrieval construct a candidate
  pool; structured domain information classifies candidate incompatibilities; and
  Pure or Mixed negatives are used to fine-tune a standard dual encoder.}
  \label{fig:pipeline}
\end{figure*}

In this section, we present \cahnm, short for \emph{Constraint-Aware Hard Negative Mining}. The framework combines conventional retrieval hardness with structured-domain incompatibility during negative selection and operationalizes Equation~\ref{eq:constraint-negative}. It proceeds in three main phases: candidate construction, structured incompatibility classification, and negative selection for retriever training. These phases are implemented through retrieval, positive removal, target-context construction, constraint classification, negative selection, training-example materialization, and dual-encoder fine-tuning. Figure~\ref{fig:pipeline} summarizes the complete pipeline.

\subsection{Candidate Construction}
\vspace{0mm}

For each query $q$, \cahnm combines lexical and dense retrieval:
\begin{equation}
  C(q)=
  \operatorname{TopK}_{\mathrm{BM25}}(q)
  \cup
  \operatorname{TopK}_{\mathrm{Dense}}(q).
  \label{eq:candidate-pool}
\end{equation}
Membership in $C(q)$ operationalizes retrieval hardness: a candidate is eligible
for further analysis when it appears among the top-$K$ results of at least one
retriever.

Known positives are removed before constraint classification:
\begin{equation}
  C^{-}(q)=C(q)\setminus P(q).
  \label{eq:positive-removal}
\end{equation}
Because incomplete judgments can leave relevant or ambiguous documents in
$C^{-}(q)$, the subsequent rules also remove candidates for which target evidence
or conflicting evidence is detected.

\paragraph{Structured Model Instantiation}: The model in Equation~\ref{eq:domain-model} is instantiated from the structure
available in each collection. Depending on the dataset, this includes
prerequisite, hierarchical, related-concept, scope, or level information.
Explicit document--concept links are used when available. Otherwise,
document--concept associations are approximated by normalized matching over
concept labels and aliases.
The structured model is used only during offline negative construction. Graph
nodes, relation types, concept identifiers, and violation categories are not
supplied as input features to the dense encoder.

\subsection{Target Context and Constraint Classification}
%\vspace{0mm}

For each query $q$, \cahnm identifies a target concept
$c_q\in\mathcal{C}$. An explicit concept identifier is used when available;
otherwise the target is obtained by normalized matching against labels and
aliases in $\mathcal{A}$.

Around the target concept, the miner constructs a local typed context:
\begin{equation}
  \mathcal{G}_q=
  \{c_q\}\cup
  \bigcup_{r\in\mathcal{R}_q}
  \operatorname{Nbr}_{r}(c_q),
  \label{eq:local-graph}
\end{equation}
where $\mathcal{R}_q\subseteq\mathcal{R}$ contains the relation types available
around $c_q$. The framework therefore does not require one fixed ontology
schema.

Each candidate $d\in C^{-}(q)$ is evaluated by a deterministic classifier:
\begin{equation}
  J(q,d,\mathcal{G}_q)=
  \bigl(\ell(q,d),\widehat{\phi}(q,d),e(q,d)\bigr),
  \label{eq:classifier}
\end{equation}
where $\ell(q,d)$ is the operational decision,
$\widehat{\phi}(q,d)$ is the detected violation set, and $e(q,d)$ records the
textual or structured evidence used by the rule.

\begin{table}[h]
  \caption{Deterministic constraint-classification rules.}
  \label{tab:rules}
  \centering
  \scriptsize
  \setlength{\tabcolsep}{2.5pt}
  \begin{tabularx}{\textwidth}{@{}p{.30\textwidth}X@{}}
    \toprule
    Classification & Detection condition \\
    \midrule
    Target match &
    Target evidence occurs and no incompatibility is detected \\
    Relational mismatch &
    A sibling, prerequisite, or postrequisite concept occurs without sufficient
    target evidence \\
    Scope or level mismatch &
    A broader or narrower concept, or an incompatible level, occurs without
    sufficient target evidence \\
    Target-context mismatch &
    A candidate associated with a related context lacks sufficient evidence for
    the target concept under the implemented matching rules \\
    Ambiguous &
    Target evidence and incompatibility evidence occur together \\
    No constraint evidence &
    Neither sufficient target evidence nor an explicit structured
    incompatibility is detected \\
    \bottomrule
  \end{tabularx}
\end{table}

Relational, scope/level, and target-context mismatches define the
constraint-aware negative pool:
\begin{equation}
  d\in\widehat{\mathcal{N}}_{\mathcal{O}}(q)
  \Longleftrightarrow
  \widehat{\phi}(q,d)\neq\emptyset
  \ \land\
  \ell(q,d)=\textsc{Mismatch}.
  \label{eq:operational-hard}
\end{equation}
Candidates classified as target matches, ambiguous, or without explicit
constraint evidence are not retained in this pool.

\subsection{Negative Selection, Controls, and Training}
\vspace{0mm}

We evaluate two uses of the constraint-aware pool. In \textit{Pure
	constraint-aware selection}, \cahnpure groups eligible candidates by their
primary violation type and selects them in round-robin order across non-empty
groups, while preserving retrieval order within each group. At most four
explicit negatives are retained per query, limiting the influence of any single
frequent violation category. As a control, \textit{retrieval-rank-matched
	selection} uses the same retrieved candidate pool but does not consult
$\mathcal{O}$. For each constraint-aware negative at retrieval position $r$,
the control searches candidates in increasing absolute rank distance from $r$
and selects the first eligible non-positive candidate, with ties resolved
according to the original retrieval order. This provides a comparison at
approximately matched retrieval positions without using structured
incompatibility during selection.

\paragraph{Mixed negative pools:}
The mixed strategy combines conventional and constraint-aware sources:
\begin{equation}
  \mathcal{N}_{\mathrm{mix}}(q)=
  \mathcal{N}_{\mathrm{rand}}(q)
  \cup
  \mathcal{N}_{\mathrm{BM25}}(q)
  \cup
  \mathcal{N}_{\mathrm{dense}}(q)
  \cup
  \widehat{\mathcal{N}}_{\mathcal{O}}(q).
  \label{eq:mixed}
\end{equation}
Duplicates are removed before applying the four-negative budget. The
source-specific pools are processed with the same deterministic allocation
procedure for \cahnmix and matched-mixed; the latter differs only by replacing
the constraint-aware source with rank-matched candidates. This comparison
therefore evaluates the constraint-aware component within the same mixed-pool
construction.

\paragraph{Training examples and fine-tuning:}
Selected examples are materialized as
\begin{equation}
  \bigl(q,d^{+},\mathcal{N}(q)\bigr),
  \label{eq:training-instance}
\end{equation}
where $d^{+}\in P(q)$ is a positive document and $\mathcal{N}(q)$ is the
negative set produced by the selected strategy. The examples fine-tune a
standard dual encoder using either triplet loss or cached multiple-negatives
ranking loss. Structured information influences training only
through negative-set composition; it is not used as an encoder input during
training or inference.

\section{Experimental Setup}
\label{sec:setup}

The experiments follow the three research questions introduced in
Section~\ref{sec:introduction}: RQ1 compares \cahnpure with rank-matched
selection, RQ2 evaluates \cahnmix against matched-mixed, and RQ3 examines
structural variants. The following subsections describe the experimental setup.

\subsection{Datasets}
\vspace{0mm}

We convert MOOCCubeX and Course-Skill Atlas into query--document retrieval tasks,
with concepts or skills as queries and learning resources or curriculum records
as documents. Relevance is derived from source concept links.

\begin{table}[t]
  \caption{Processed collection statistics. Query and qrel counts are reported
  as train/development/test.}
  \label{tab:data}
  \centering
  \scriptsize
  \begin{tabular}{lrrrr}
    \toprule
    Collection & Documents & Queries & Qrels & Relations \\
    \midrule
    MOOCCubeX
    & 3,781
    & 3,500/500/1,000
    & 10,241/1,453/3,025
    & \texttt{requires} \\
    Course-Skill Atlas
    & 281,153
    & 227/32/65
    & 22,700/3,200/6,500
    & \texttt{is\_a}, \texttt{related} \\
    \bottomrule
  \end{tabular}
\end{table}

Splits are query-level, with no query or target overlap across train,
development, and test partitions. Because Course-Skill Atlas has more relevant
documents per query than MOOCCubeX, results are interpreted separately by
collection. In Course-Skill Atlas, each DWA query is linked to
institution--field--year profiles whose source top-10 DWA list contains that
activity. After deduplication, up to 100 profiles define the binary
judged-positive set $P(q)$; profiles beyond this cutoff remain unjudged rather
than known negatives.

\subsection{Compared Methods}
\vspace{0mm}

The principal experiment compares four negative strategies:
\begin{itemize}
  \item \textbf{DPR-Random}: uniformly sampled non-positive documents
  ~\cite{karpukhin2020dense};
  \item \textbf{DenseNeg}: the highest-ranked non-positive documents returned by
  the initial dense retriever~\cite{xiong2020approximate,zhan2021optimizing};
  \item \textbf{Rank-matched}: non-positive candidates selected from retrieval
  positions close to those of \cahnpure, without using structured constraints;
  \item \textbf{\cahnpure}: constraint-aware negatives selected by the procedure
  in Section~\ref{sec:method}.
\end{itemize}

For \textbf{RQ2}, \cahnmix is compared with the conventional negative
sources and with \textbf{matched-mixed}, which replaces its constraint-aware
component with rank-matched candidates. For \textbf{RQ3}, we compare
\cahnpure with three structural variants. \textbf{Label-only} retains
label/alias evidence while omitting typed relation context;
\textbf{shuffled-graph} replaces the original relation neighborhood with a
shuffled one; and \textbf{no-structure} disables relation-derived context and
uses only the remaining non-structural evidence available to the classifier.
Because these variants can also change candidate eligibility and negative-pool
composition, they are analyzed as end-to-end sensitivity variants rather than
relation-level interventions.

The principal experiment uses two datasets, two encoders, two losses, four
negative strategies, and ten seeds. We fine-tune BGE
~\cite{xiao2023cpack} and E5~\cite{wang2022textembeddings} with triplet
loss~\cite{schroff2015facenet} or cached MNRL using Sentence Transformers
~\cite{reimers2019sentence}. Training uses three epochs, batch size 32,
learning rate $2\times10^{-5}$, 10\% warmup, 128-token inputs, and four
explicit negatives per query. Candidate mining uses $K=100$ and seed 13;
training uses ten fixed seeds. RQ2 and RQ3 use BGE with cached MNRL.
Source code and reproduction scripts are available online.\footnote{
	\url{https://github.com/lengocluyen/ca-hnm}}

\subsection{Evaluation and Statistical Analysis}
\vspace{0mm}

We report \ndcg, \mrr, \mapscore, and \recall. Both collections use binary qrels.
\ndcg summarizes top-ranked quality, \mrr emphasizes the first relevant result,
\mapscore summarizes precision over relevant documents, and \recall measures
coverage in the top 100 ranks~\cite{jarvelin2002cumulated}.

For \textbf{RQ1}, the paired inferential analysis focuses on the
\cahnpure versus rank-matched difference in \ndcg. Differences are paired by
training seed, and hierarchical 95\% bootstrap confidence intervals are obtained
by resampling seeds and queries. The remaining metrics provide complementary
descriptive evidence about ranking quality and retrieval coverage. For
\textbf{RQ2}, we report the mean \ndcg difference between \cahnmix and
matched-mixed in the BGE/cached-MNRL setting. \textbf{RQ3} is interpreted as an
end-to-end sensitivity analysis because its structural variants can change
candidate eligibility as well as the available structured information. Because
Course-Skill Atlas contains 32 development queries, interval width is considered
together with the mean difference.

\section{Results and Analysis}
\label{sec:results}

The results are organized according to the three research questions introduced
in Section~\ref{sec:introduction}. We first examine the rank-matched comparison,
then the mixed negative pool, and finally the structural sensitivity analysis.

\subsection{RQ1: Comparison with Rank-Matched Selection}
\vspace{0mm}
\label{sec:rq1}

Table~\ref{tab:core} reports the full comparison across datasets, encoders,
losses, and negative strategies.
Absolute performance depends strongly on the training configuration. On
Course-Skill Atlas, \cahnpure has the highest mean score on all four metrics for
E5 with triplet loss. DPR-Random leads the BGE/triplet and E5/cached-MNRL
settings on most or all metrics, while DenseNeg has the highest \ndcg and \mrr
for BGE/cached MNRL. On MOOCCubeX, DPR-Random or DenseNeg has the highest
absolute mean score in every encoder--loss block.

\begin{table}[!t]
	\caption{Development retrieval performance averaged over ten training seeds. Bold marks the best strategy for each encoder--loss--metric setting.}
	\label{tab:core}
	\centering
	\scriptsize
	\setlength{\tabcolsep}{2.6pt}
	\begin{tabular}{lllrrrr}
		\toprule
		Encoder & Loss & Strategy & NDCG@10 & MRR@10 & MAP@100 & R@100 \\
		\midrule
		\multicolumn{7}{c}{\textbf{Course-Skill Atlas}} \\
		\midrule
		BGE & Cached MNRL & DPR-Random & 0.0608 & 0.0853 & \textbf{0.0177} & \textbf{0.0775} \\
		&  & DenseNeg & \textbf{0.0901} & \textbf{0.1911} & 0.0153 & 0.0696 \\
		&  & Rank-matched & 0.0379 & 0.0598 & 0.0074 & 0.0425 \\
		&  & \textsc{CA-HNM}-Pure & 0.0311 & 0.0581 & 0.0066 & 0.0384 \\
		\addlinespace[1pt]
		& Triplet & DPR-Random & \textbf{0.2283} & \textbf{0.3203} & \textbf{0.0713} & \textbf{0.1550} \\
		&  & DenseNeg & 0.0447 & 0.0900 & 0.0074 & 0.0382 \\
		&  & Rank-matched & 0.0569 & 0.1043 & 0.0109 & 0.0500 \\
		&  & \textsc{CA-HNM}-Pure & 0.0585 & 0.1052 & 0.0115 & 0.0485 \\
		\addlinespace[1pt]
		E5 & Cached MNRL & DPR-Random & \textbf{0.1777} & \textbf{0.2422} & \textbf{0.0477} & \textbf{0.1441} \\
		&  & DenseNeg & 0.0797 & 0.1447 & 0.0137 & 0.0686 \\
		&  & Rank-matched & 0.0764 & 0.1078 & 0.0180 & 0.0761 \\
		&  & \textsc{CA-HNM}-Pure & 0.0738 & 0.0935 & 0.0190 & 0.0765 \\
		\addlinespace[1pt]
		& Triplet & DPR-Random & 0.0705 & 0.0935 & 0.0146 & 0.0581 \\
		&  & DenseNeg & 0.0158 & 0.0423 & 0.0011 & 0.0082 \\
		&  & Rank-matched & 0.2036 & 0.3513 & 0.0364 & 0.0981 \\
		&  & \textsc{CA-HNM}-Pure & \textbf{0.2947} & \textbf{0.4777} & \textbf{0.0574} & \textbf{0.1340} \\
		\midrule
		\multicolumn{7}{c}{\textbf{MOOCCubeX}} \\
		\midrule
		BGE & Cached MNRL & DPR-Random & \textbf{0.3280} & \textbf{0.3031} & \textbf{0.2980} & \textbf{0.8808} \\
		&  & DenseNeg & 0.3020 & 0.2869 & 0.2700 & 0.7164 \\
		&  & Rank-matched & 0.1547 & 0.1405 & 0.1373 & 0.5219 \\
		&  & \textsc{CA-HNM}-Pure & 0.1724 & 0.1596 & 0.1530 & 0.5657 \\
		\addlinespace[1pt]
		& Triplet & DPR-Random & 0.3352 & 0.3339 & 0.2698 & 0.9264 \\
		&  & DenseNeg & \textbf{0.3511} & \textbf{0.3505} & \textbf{0.2710} & \textbf{0.9288} \\
		&  & Rank-matched & 0.1653 & 0.1755 & 0.1294 & 0.5684 \\
		&  & \textsc{CA-HNM}-Pure & 0.2472 & 0.2592 & 0.1880 & 0.5918 \\
		\addlinespace[1pt]
		E5 & Cached MNRL & DPR-Random & \textbf{0.3662} & \textbf{0.3391} & \textbf{0.3314} & \textbf{0.9099} \\
		&  & DenseNeg & 0.2649 & 0.2516 & 0.2375 & 0.6317 \\
		&  & Rank-matched & 0.1750 & 0.1632 & 0.1583 & 0.5625 \\
		&  & \textsc{CA-HNM}-Pure & 0.1785 & 0.1682 & 0.1635 & 0.5731 \\
		\addlinespace[1pt]
		& Triplet & DPR-Random & \textbf{0.3930} & \textbf{0.3963} & \textbf{0.2982} & \textbf{0.9158} \\
		&  & DenseNeg & 0.1755 & 0.1946 & 0.1307 & 0.5494 \\
		&  & Rank-matched & 0.0694 & 0.0675 & 0.0536 & 0.3629 \\
		&  & \textsc{CA-HNM}-Pure & 0.0962 & 0.1138 & 0.0727 & 0.3756 \\
		\bottomrule
	\end{tabular}
\end{table}

The comparison with rank-matched selection addresses a different question:
whether structured selection changes the training signal when retrieval position
is approximately controlled. \cahnpure has higher mean \ndcg in six of the eight
dataset--encoder--loss configurations. Under triplet loss, the mean difference is
positive in all four configurations; under cached MNRL it is positive in two of
four.

Four hierarchical 95\% bootstrap intervals are entirely above zero:
Course-Skill Atlas with E5/triplet
($\Delta=+0.0911$, $[0.0617,\,0.1228]$),
MOOCCubeX with BGE/cached MNRL
($\Delta=+0.0177$, $[0.0057,\,0.0298]$),
MOOCCubeX with BGE/triplet
($\Delta=+0.0819$, $[0.0589,\,0.1042]$), and
MOOCCubeX with E5/triplet
($\Delta=+0.0268$, $[0.0079,\,0.0495]$).
The other four intervals span zero.

\begin{figure}[H]
  \centering
  \includegraphics[width=.99\textwidth]{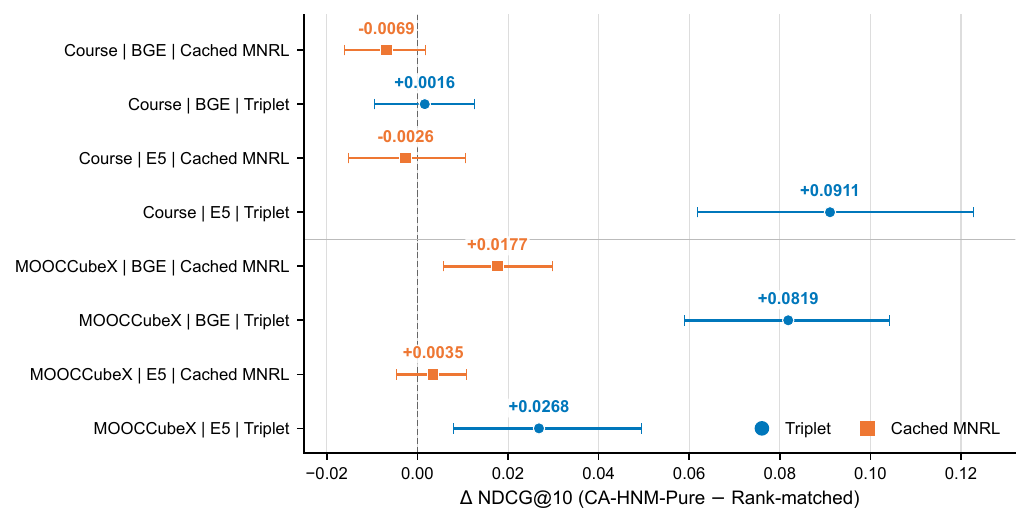}
  \caption{\cahnpure minus rank-matched development \ndcg. Points show paired
  ten-seed mean differences and bars show hierarchical 95\% bootstrap confidence
  intervals obtained by resampling seeds and queries.}
  \label{fig:effect-forest}
\end{figure}

The pattern is therefore loss-dependent rather than uniform across settings.
Triplet loss contains three of the four intervals that are entirely above zero,
whereas the cached-MNRL differences are smaller except for MOOCCubeX with BGE.

\subsection{RQ2: Mixed Negative Pools}
\vspace{0mm}
\label{sec:rq2}

Table~\ref{tab:hybrid} evaluates \cahnmix under BGE with cached MNRL.
Matched-mixed replaces the constraint-aware component with rank-matched
candidates while retaining the other negative sources.

\begin{table}[H]
	\caption{Main comparison and component ablation of \textsc{CA-HNM}-Mixed with BGE and cached MNRL, averaged over ten training seeds. Matched-mixed replaces its constraint-aware component with rank-matched negatives. NDCG and MRR use cutoff 10; MAP and recall use cutoff 100.}
	\label{tab:hybrid}
	\centering
	\scriptsize
	\setlength{\tabcolsep}{2.6pt}
	\begin{tabular}{lrrrrrrrr}
		\toprule
		& \multicolumn{4}{c}{Course-Skill Atlas} & \multicolumn{4}{c}{MOOCCubeX} \\
		\cmidrule(lr){2-5}\cmidrule(lr){6-9}
		Strategy & NDCG & MRR & MAP & R@100 & NDCG & MRR & MAP & R@100 \\
		\midrule
		DPR-Random & 0.0608 & 0.0853 & 0.0177 & 0.0775 & \textbf{0.3280} & \textbf{0.3031} & \textbf{0.2980} & \textbf{0.8808} \\
		DenseNeg & 0.0901 & \textbf{0.1911} & 0.0153 & 0.0696 & 0.3020 & 0.2869 & 0.2700 & 0.7164 \\
		\textsc{CA-HNM}-Mixed & \textbf{0.1120} & 0.1597 & \textbf{0.0238} & \textbf{0.0818} & 0.3031 & 0.2848 & 0.2747 & 0.8090 \\
		Matched-mixed & 0.1082 & 0.1633 & 0.0214 & 0.0777 & 0.2955 & 0.2759 & 0.2676 & 0.8041 \\
		\bottomrule
	\end{tabular}
\end{table}

On Course-Skill Atlas, \cahnmix has the highest mean \ndcg, \mapscore, and
\recall, while DenseNeg has the highest mean \mrr. On MOOCCubeX, DPR-Random has
the highest mean score on all four metrics; \cahnmix is above DenseNeg on
\ndcg, \mapscore, and \recall.

Relative to matched-mixed, \cahnmix changes mean \ndcg by $+0.0038$ on
Course-Skill Atlas and $+0.0076$ on MOOCCubeX. The remaining mean differences
favor \cahnmix on three of four Course-Skill Atlas metrics and all four
MOOCCubeX metrics. The \ndcg differences between the two mixed-pool variants
are small in this BGE/cached-MNRL experiment.

\subsection{RQ3: Structural Sensitivity}
\vspace{0mm}
\label{sec:rq3}

Table~\ref{tab:abl} compares \cahnpure with label-only, shuffled-graph, and
no-structure variants. The structural variants behave differently across the two collections. On
Course-Skill Atlas, the shuffled-graph variant has the highest mean value on all
four metrics, and no-structure is also above \cahnpure on most metrics. On
MOOCCubeX, the four variants are much closer, with no-structure having the
highest mean value on all four metrics.

\begin{table}[H]
	\caption{Structural ablation of \textsc{CA-HNM}-Pure with BGE and cached MNRL, averaged over ten training seeds.}
	\label{tab:abl}
	\centering
	\scriptsize
	\setlength{\tabcolsep}{2.6pt}
	\begin{tabular}{lrrrrrrrr}
		\toprule
		& \multicolumn{4}{c}{Course-Skill Atlas} & \multicolumn{4}{c}{MOOCCubeX} \\
		\cmidrule(lr){2-5}\cmidrule(lr){6-9}
		Strategy & NDCG & MRR & MAP & R@100 & NDCG & MRR & MAP & R@100 \\
		\midrule
		\textsc{CA-HNM}-Pure & 0.0311 & 0.0581 & 0.0066 & 0.0384 & 0.1724 & 0.1596 & 0.1530 & 0.5657 \\
		Label-only & 0.0263 & 0.0417 & 0.0067 & 0.0389 & 0.1733 & 0.1581 & 0.1528 & 0.5617 \\
		Shuffled graph & \textbf{0.1174} & \textbf{0.1895} & \textbf{0.0220} & \textbf{0.0745} & 0.1744 & 0.1614 & 0.1540 & 0.5666 \\
		No structure & 0.0950 & 0.1669 & 0.0166 & 0.0693 & \textbf{0.1835} & \textbf{0.1690} & \textbf{0.1629} & \textbf{0.5768} \\
		\bottomrule
	\end{tabular}
\end{table}

These variants modify the information available to the classifier and can also
change candidate eligibility. Their scores therefore reflect both the structural
change and the resulting negative-pool composition. A relation-level analysis
would require candidate identities, retrieval positions, and pool sizes to remain
fixed while relation semantics are modified. The present experiment therefore
characterizes end-to-end sensitivity of the mining procedure.

\FloatBarrier

\section{Discussion}
\label{sec:discussion}

The rank-matched comparison distinguishes the effect of negative
source from differences in retrieval position. The results indicate that the
difference between constraint-aware and rank-matched selection depends on the
training configuration, with a more consistent positive pattern under triplet
loss than under cached MNRL. This pattern may reflect how the two objectives use
explicit negatives. Triplet loss contrasts each query--positive pair with the
selected negative, whereas cached MNRL distributes competition across explicit,
cached, and in-batch examples. The present experiments do not estimate the
interaction between loss function and negative selection, so this interpretation
remains to be tested directly.

The mixed-pool experiment separates the performance of the complete
negative pool from the contribution of its constraint-aware component. The mean
\ndcg differences between \cahnmix and matched-mixed are small on both
collections under BGE with cached MNRL. These results indicate that the effect
of replacing one negative source can depend on the composition of the remaining
training pool.

The structural variants show that retrieval performance is sensitive
to changes in the information available to the constraint classifier. On
Course-Skill Atlas, the shuffled-graph and no-structure variants obtain higher
mean scores than the full structured version, whereas the differences are
smaller on MOOCCubeX. These comparisons cannot be attributed solely to relation
semantics because modifying the structure can also change candidate eligibility
and negative-pool composition. A relation-level analysis would require the
candidate identities, retrieval positions, and pool sizes to remain fixed while
only the structural relations are modified.

The rule-based classifier also records the basis of each negative-selection
decision. Selected candidates are associated with a violation category and the
textual or structured evidence that triggered the corresponding rule. These
records provide a means to inspect the composition of the mined negative pool
and the relative frequency of different rule types. They should be interpreted
as operational mining annotations rather than independently validated relevance
judgments.

The scope of the study is limited in several respects. Both datasets are from
educational domains and contain explicit structured metadata. All reported
results use development partitions, and Course-Skill Atlas contains only 32
development queries. The constraint classifier relies on source labels, aliases,
lexical evidence, and manually specified rules. Rank matching approximates
retrieval position but does not control all candidate characteristics, and the
mixed-pool and structural experiments are restricted to BGE with cached MNRL.
Further evaluation should therefore include structured domains, and structural controls that preserve candidate identities
and pool composition.

\section{Conclusion}
\label{sec:conclusion}

We presented \cahnm, a hard-negative mining framework that combines
retrieval-based candidate generation with structured incompatibility rules while
retaining a standard dual encoder at inference time. Across two educational
retrieval collections, two encoders, two losses, and ten seeds, \cahnpure has
higher mean \ndcg than rank-matched selection in six of eight configurations,
with four hierarchical 95\% bootstrap intervals entirely above zero. The pattern
is more consistent under triplet loss than under cached MNRL. In the
BGE/cached-MNRL mixed experiment, \cahnmix is close to matched-mixed in
\ndcg, while the structural variants show that changing the available structure
can substantially alter the resulting negative pool. Future work will extend the analysis to other structured domains, and use
fixed-candidate structural perturbations to separate relation semantics from
candidate-composition effects.

\section*{Acknowledgments}
We thank the Ikigai consortium led by the association Games for Citizens, the company Gamaizer, and the FORTEIM project (winner of the AMI CMA France 2030 call for projects), for their support and collaboration. Their contributions have provided significant added value to the completion of this research.

\bibliographystyle{splncs04}
\bibliography{references}

\end{document}